\documentclass[12pt]{article} 

\usepackage{latexsym} 
\usepackage{amssymb}  
\usepackage{epsfig}       

\newcommand{\beq}{\begin{equation}}
\newcommand{\eeq}{\end{equation}}
\newcommand{\ba}{\begin{array}}
\newcommand{\bea}{\begin{eqnarray}}
\newcommand{\ea}{\end{array}}
\newcommand{\eea}{\end{eqnarray}}

\newcommand\comment[1]{ \hbox{[{\it Comment suppressed here.}\/]} }
\newcommand\hide[1]{}

\newcommand{\skipover}[1]{}

\def\ZZ{Z \kern -.43em Z}
\newcommand{\nnn} {\nonumber \vspace{.2cm} \\ }

\makeatletter 

\def\appendix{\par                              
    \setcounter{section}{0}                     
    \setcounter{subsection}{0}
    \renewcommand{\theequation}{\Alph{section}.\arabic{equation}}
    \renewcommand{\thesection}{Appendix \Alph{section}
                \setcounter{equation}{0}  } 
}

\def\applabel#1{\@bsphack
  \protected@write\@auxout{}%
         {\string\newlabel{#1}{{\Alph{section}}{\thepage}}}%
  \@esphack}

\def\section{
\setcounter{equation}{0}        
\@startsection {section}{1}{\z@}{-3.5ex plus -1ex minus 
 -.2ex}{2.3ex plus .2ex}{\large\bf}}
\renewcommand{\theequation}{\arabic{section}.\arabic{equation}}

\def\subsection{\@startsection{subsection}{2}{\z@}{-3.25ex plus -1ex minus 
 -.2ex}{1.5ex plus .2ex}{\normalsize\bf}}

\def\subsubsection{\@startsection{subsubsection}{3}{\z@}{-3.25ex plus
 -1ex minus -.2ex}{1.5ex plus .2ex}{\normalsize}}

\makeatother   

\newsavebox{\eqlabel}

\makeatletter  
\newlength{\numblen}
\newsavebox{\eqnumb}
\def\@eqnnum{\savebox{\eqnumb}{\rm (\theequation)}%
\settowidth{\numblen}{\usebox{\eqnumb}}%
\makebox[\numblen][l]{\usebox{\eqnumb}~~~\usebox{\eqlabel}}}
\makeatother   

\newenvironment{equationwithlabel}[1]{ %
  \begin{equation}\label{#1} }{\end{equation}} 
\newcommand{\beql}[1]{\begin{equationwithlabel}{#1}}
\newcommand{\eeql}{\end{equationwithlabel}}

\begin{document}

\title{\bf Quark description of nuclear matter
\\}

\author{J\"urgen Berges \\[1.ex]
{\normalsize Institut f{\"u}r Theoretische Physik}\\
{\normalsize Philosophenweg 16, 69120 Heidelberg, Germany}
}

\newcommand{\preprintno}{
  \normalsize (HD-THEP-00-60)
}

\date{ \preprintno}

\begin{titlepage}
\maketitle
\def\thepage{}          

\begin{abstract}
We discuss the role of an adjoint chiral condensate for 
color superconducting quark matter. Its presence leads 
to color-flavor locking in two-flavor quark matter. 
Color is broken completely as well as chiral symmetry
in the two-flavor theory with coexisting adjoint quark-antiquark
and antitriplet quark-quark condensates.
The qualitative properties of this phase match the properties of 
ordinary nuclear matter without strange baryons. This complements 
earlier proposals by Sch{\"a}fer and Wilczek for a quark 
description of hadronic phases.
We show for a class of models with effective four-fermion 
interactions that adjoint chiral and diquark condensates do not 
compete, in the sense that simultaneous condensation occurs
for sufficiently strong interactions in the adjoint chiral
channel.
\end{abstract}

\end{titlepage}

\renewcommand{\thepage}{\arabic{page}}



\section{Introduction}

Important progress in understanding the behavior of hadronic matter at 
high baryon number density has been achieved recently \cite{review}. 
In QCD with three colors and three flavors of quarks the sufficiently
high density phase exhibits chiral symmetry breaking and confinement
which can be studied in weak coupling. The color superconducting
ground state at high density is characterized by the phenomenon of
color-flavor locking \cite{ARWCFL}. As a consequence all the elementary
excitations carry integral electric charges and for the most part
match the quantum numbers of hypernuclear matter 
\cite{SWCONT,ABR,SWCONT2}. At densities
below the threshold for strangeness the situation seems to change
dramatically: The color superconducting ground state for two quark
flavors no longer breaks chiral symmetry and color symmetry is restored 
partially \cite{Barrois,BailinLove,ARW,Rapp,BR}. 

The importance of ``color symmetry breaking'' in the vacuum for
QCD with three flavors and three colors of quarks has been 
recently pointed out in \cite{CWCFL}. According to \cite{CWINST} 
instanton induced interactions result in a nonvanishing color
octet quark-antiquark condensate. The new vacuum exhibits color-flavor
locking, confinement admits an equivalent description in a Higgs picture
and the excitations match properties of the
low momentum meson and baryon degrees of freedom. A consistent picture
of strong interactions at long distances emerges \cite{CWCFL}. 

In this letter we discuss the role of an adjoint quark-antiquark 
condensate at high density. We concentrate on the two-flavor limit of 
the octet condensate proposed in \cite{CWCFL,CWINST}. This limit exhibits 
two-flavor color-flavor locking which leaves a global diagonal
$SU(2)_{{\rm color} +V}$ invariant. A more general group theoretical discussion
of color-flavor locking for two quark flavors will be presented in 
\cite{BWW}. While for three flavors
condensation in the adjoint quark-antiquark channel does not change
the present understanding of the high density ground state, for two 
quark flavors the picture changes dramatically. For simultaneous condensation
in the adjoint chiral and the color-antitriplet diquark channels 
color symmetry is broken completely as well as chiral symmetry.
The qualitative properties of this phase match the properties of 
nuclear matter without strange baryons, extending 
earlier proposals \cite{SWCONT2,SWCONT} for a quark 
description of hadronic phases. 

We discuss possible mechanisms 
for simultaneous condensation at nonzero density in effective fermionic 
models. We show that simultaneous condensation indeed occurs  
for sufficiently strong attractive interaction in the adjoint 
quark-antiquark channel.   
Earlier investigations studying possibilities for coexisting condensates
in the standard color-singlet quark-antiquark and antitriplet diquark
channels found a strong competition and no simultaneous condensation 
\cite{BR,CD}. In this sense the adjoint chiral condensate does not 
compete with the diquark condensate.  
The result is encouraging and opens very attractive
possibilities for a quark description of ordinary nuclear matter.
We stress, however, that a strong attraction in the adjoint chiral
channel is a necessary condition which may not be realized in the
two-flavor theory and for which we cannot claim control here.

\section{The adjoint 
chiral condensate at high density}

\hspace*{0.7cm}
{\bf Three quark flavors.\ }
In QCD with three flavors of massless quarks the Cooper
pairs at high density cannot be flavor singlets, and both
flavor and color symmetries are necessarily broken. A 
global, vector-like $SU(3)$ symmetry can be retained by
the phenomenon of color-flavor locking \cite{ARWCFL}. Color-flavor
locking means that left flavor rotations and 
transposed color rotations are performed with same opposite 
angles. The same happens separately with the right flavor
rotations and color rotations. Since color is a vectorial
symmetry this leaves a global diagonal $SU(3)_{{\rm color}+V}$
and breaks chiral symmetry. In particular, all gluons acquire
a mass by the Higgs mechanism. 

The notion of color-flavor locking
is not restricted to Cooper pairs of quarks but can also be
applied to quark--antiquark condensates. This has been pointed out 
in \cite{CWCFL} where a color-flavor locking quark-antiquark condensate in the
octet representation is proposed in the QCD vacuum. 
For three massless flavors the 
color-flavor locking quark--antiquark ($\chi^{\rm (3)}$) and
quark--quark ($\Delta^{\rm (3)}$) condensates  
are (omitting Dirac structure) 
\bea
\chi^{\rm (3)} &\sim&
\Big\langle \bar{\psi}^{\alpha}_a \sum\limits_{s=1}^8
\left( \lambda_s \right)_{ab} \left( \lambda_s \right)^{ \beta\alpha} 
\psi^{\beta}_b \Big\rangle 
\nnn
&\equiv& 
\Big\langle \bar{\psi}^{\alpha}_a 
\Big(2 \delta_a^\alpha \delta_b^\beta 
- \frac{2}{3} \delta_{ab} \delta^{\alpha\beta}\Big)
\psi^{\beta}_b \Big\rangle   \,\, ,
\label{chi3CFL}\\[0.3cm]
\Delta^{\rm (3)} &\sim&
\Big\langle \psi^{\alpha}_a \sum\limits_{s=1}^8
\left( \lambda_s \right)_{ab} \left( \lambda_s \right)^{\beta\alpha } 
\psi^{\beta}_b \Big\rangle
\nnn
&\equiv&
\Big\langle \psi^{\alpha}_a \sum\limits_{j=1}^3
\left( \lambda^{(A)}_j \right)_{ab} 
\left( \lambda^{(A)}_j \right)^{\beta\alpha} 
\psi^{\beta}_b \Big\rangle_{\bf \bar{3}_c}
+ \Big\langle \psi^{\alpha}_a \sum\limits_{i=1}^5
\left( \lambda^{(S)}_i \right)_{ab} 
\left( \lambda^{(S)}_i \right)^{\beta\alpha } 
\psi^{\beta}_b \Big\rangle_{\bf 6_c}
\nnn
&\equiv& 
\Big\langle {\psi}^{\alpha}_a 
\Big(\delta_a^\alpha \delta_b^\beta 
- \delta_{a}^\beta \delta_b^{\alpha} \Big)
\psi^{\beta}_b \Big\rangle_{\bf \bar{3}_c}
+ \Big\langle {\psi}^{\alpha}_a 
\Big(\delta_a^\beta \delta_b^\alpha 
+ \delta_{a}^\alpha \delta_b^{\beta} 
- \frac{2}{3} \delta_{ab} \delta^{\alpha\beta} \Big)
\psi^{\beta}_b \Big\rangle_{\bf 6_c}
\label{di3CFL}\nonumber \\
\eea
Here $\lambda^s$ denote the eight Gell-Mann matrices for flavor
($a,b=1,\ldots,8$) and color ($\alpha,\beta = 1,\ldots,8$). 
We have separated for the diquark condensate the 
color-antitriplet (${\bf \bar{3}_c}$) and the color-symmetric contribution 
(${\bf 6_c}$). Since the antitriplet and the sextet condensate leave 
the same symmetries invariant any relative weight differing from one
between the two contributions in (\ref{di3CFL}) 
leads to color-flavor locking. The ${\bf \bar{3}_c}$- and the 
${\bf 6_c}$-condensate can mix, and in general a nonzero 
${\bf \bar{3}_c}$-condensate induces a non-vanishing 
${\bf 6_c}$ or vice versa. 
At high density the color--${\bf 6}$ contribution to the color-flavor 
locked state is found to be small \cite{ARWCFL}. 

In the color superconducting high density phase with $\Delta^{(3)} \not = 0 $
a nonzero adjoint chiral condensate $\chi^{(3)}$ does not break any
new symmetries\footnote{In a theory with $U(1)_V \times U(1)_A$ symmetry
the diquark condensate leaves a residual $\ZZ^L_2 \times \ZZ^R_2$ symmetry
invariant \cite{ARWCFL}. If this were a valid symmetry the 
adjoint chiral condensate would vanish. Instanton-induced interactions
break the $\ZZ^L_2$ and in presence of the diquark condensate and the
't Hooft vertex one can have an adjoint chiral condensate.} and a nonzero
$\Delta^{(3)}$ induces a non-vanishing value for $\chi^{(3)}$ 
\cite{CWCFL}. At high density the color-octet condensate is, therefore,
present.\\

{\it Integer charges and quark description of hypernuclear matter.\ }
The color-flavor locking condensates (\ref{chi3CFL}) and (\ref{di3CFL})
break chiral symmetry and produce gaps for all three flavors and
all three colors. All gluons acquire a mass according to the Higgs
phenomenon. They also break the $U(1)$ of electromagnetism but the
condensates are
invariant under a combination $\tilde{Q}=Q-Q_c$ 
of electric charge $Q$ and abelian
color charge $Q_c=\frac{1}{2}\lambda_3+\frac{1}{2 \sqrt{3}} \lambda_8$
\cite{ARWCFL,CWCFL}. Baryon number is broken only by the diquark condensate. 

The dressed elementary excitations in presence of the color-flavor
locking condensates carry integer electric charge with respect
to $\tilde{Q}$ and for the most part match the expected features of 
confined hadronic matter. The gluons match the octet of vector mesons, 
the quark octet matches the baryon octet and an octet of collective
modes due to chiral symmetry breaking is associated with the
pseudoscalar octet \cite{SWCONT,CWCFL}. At high densities, where diquark
condensation breaks baryon number the quark matter description   
matches superfluid hypernuclear matter at densities where strange baryons 
are present \cite{SWCONT,ABR,SWCONT2}. 
Since there is no difference in 
symmetries between the quark and the hypernuclear matter phase 
there is no need for a phase transition. This is {\it continuity}
as pointed out by Sch{\"a}fer and Wilczek \cite{SWCONT}.  \\

{\bf Two quark flavors.\ }
Color-flavor locking in the limit of two quark flavors can be more involved
since the number of flavor and color generators do not match as in the
three-flavor case. 
In contrast to the latter, there is in particular the possibility to form 
a flavor singlet diquark in the two-flavor theory which does not
break chiral symmetry \cite{Barrois,BailinLove,ARW,Rapp}. Color-flavor
locking may therefore not be expected in a two-flavor theory with
diquark condensates only. 

The two-flavor case can be obtained straightforwardly from the
three-flavor case by sending the strange quark mass to infinity.  
In ref.\ \cite{ABR,SWCONT2} the strange quark
mass dependence of the color-flavor locking diquark state has been discussed 
in a theory with two massless quarks and one strange quark.
For non-zero but sufficiently small strange quark mass one finds 
that diquark condensation leaves a global 
$SU(2)_{{\rm color} + V}$ subgroup invariant. This is achieved
by locking the left and right $SU(2)$-flavor symmetries to an $SU(2)$ 
subgroup of color. If the density dependent effective strange quark 
mass $M_s(\mu)$ exceeds $\simeq (2 \mu \Delta_{us})^{1/2}$, where
$\Delta_{us}$ denotes condensates which pair a strange quark with
either an up or a down quark, then an unlocking transition is observed
\cite{ABR}. For two quark flavors, in the limit of infinite
strange quark mass, no color-flavor locking and consequently no
chiral symmetry breaking is observed in the diquark sector. 
 
In refs.\
\cite{BR,CD} it has been shown that the standard color-singlet
chiral condensate competes with the diquark condensate. 
No simultaneous condensation was observed and apart from the 
small explicit breaking due to nonzero
quark masses there is no chiral symmetry breaking. 

The two-flavor theory has so far not been discussed in the
light of a possible adjoint chiral condensate. In ref.\ \cite{BWW} 
a general group theoretical discussion of two-flavor color-flavor 
locking is presented. Here we restrict ourselves to consider the
two-flavor (infinite strange quark mass) limit of the condensates 
(\ref{chi3CFL}) and (\ref{di3CFL}) which read
\bea
\chi^{\rm (2)} &\sim& 
\Big\langle \bar{\psi}^{\alpha}_a \sum\limits_{s=1}^3
\left( \tau_s \right)_{ab} \left( \lambda_s \right)^{ \beta\alpha} 
\psi^{\beta}_b \Big\rangle
\nnn
&\equiv& \Big\langle \bar{\psi}^{\alpha}_a 
\Big(2 \delta_a^\alpha \delta_b^\beta - \delta_{ab} \delta^{\alpha\beta}\Big)
\psi^{\beta}_b \Big\rangle   \qquad {\rm for\ } \alpha,\beta = 1,2 \,\, , 
\label{chicond}
\\[0.3cm]
\Delta^{\rm (2)} &\sim&
\Big\langle \psi^{\alpha}_a \sum\limits_{s=1}^3
\left( \tau_s \right)_{ab} \left( \lambda_s \right)^{\beta\alpha} 
\psi^{\beta}_b \Big\rangle
\nonumber\\[-0.2cm]
&\equiv&
\Big\langle \psi^{\alpha}_a 
\left( \tau_2 \right)_{ab} \left( \lambda_2 \right)^{\beta\alpha } 
\psi^{\beta}_b \Big\rangle_{\bf \bar{3}_c}
+ \Big\langle \psi^{\alpha}_a \sum\limits_{i=1}^2
\left( \tau^{(S)}_i \right)_{ab} \left( \lambda^{(S)}_i \right)^{\beta\alpha } 
\psi^{\beta}_b \Big\rangle_{\bf 6_c}
\nnn
&\equiv& 
\Big\langle {\psi}^{\alpha}_a 
\Big(\delta_a^\alpha \delta_b^\beta 
- \delta_{a}^\beta \delta_b^{\alpha} \Big)
\psi^{\beta}_b \Big\rangle_{\bf \bar{3}_c}
+ \Big\langle {\psi}^{\alpha}_a 
\Big(\delta_a^\beta \delta_b^\alpha 
+ \delta_{a}^\alpha \delta_b^{\beta} 
- \delta_{ab} \delta^{\alpha\beta} \Big)
\psi^{\beta}_b \Big\rangle_{\bf 6_c}
\nnn
&\sim& \Delta^{(2)}_{\bf \bar{3}_c} + \Delta^{(2)}_{\bf 6_c}
\label{delcond}
\nonumber \\
\eea
where the color indices in the last equation are
$\alpha,\beta=1,2$. Here $\tau^{(S)}_i$ are the symmetric
Pauli matrices and $\lambda^{(S)}_i$ the corresponding 
symmetric Gell-Mann matrices $\lambda_1$ and $\lambda_3$.
 
Since the adjoint quark--antiquark condensate 
$\chi^{\rm (2)}$ given in (\ref{chicond}) is a flavor triplet 
it breaks both color and flavor symmetries separately. It is 
invariant under the global diagonal $SU(2)_{{\rm color}+V}$ subgroup
which applies color and flavor transformations simultaneously.  
The condensate therefore
locks flavor-$SU(2)$ rotations to an $SU(2)$ subgroup of color and 
extends the notion of color-flavor locking to a pure two-flavor
theory. Condensation of the form (\ref{chicond}) leaves a local $U(1)_8$ 
subgroup of color unbroken. It therefore gives a mass to seven of the 
eight gluons by the Higgs mechanism \cite{BWW}.   

There is a crucial difference between the two- and the three-flavor 
case for the diquark channel as mentioned above. 
The color-antitriplet part of the two-flavor diquark condensate 
$\Delta^{(2)}_{\bf \bar{3}_c}$ given in
(\ref{delcond}) is a flavor singlet and it doesn't break 
chiral symmetry. It also leaves an $SU(2)$ subgroup of color--$SU(3)$ 
unbroken and consequently five massless gluons remain.
Therefore, for two flavors there is  
no color-flavor locking in the ${\bf \bar{3}_c}$--channel. 
In contrast, a flavor-symmetric part of the
condensate ($\Delta^{(2)}_{\bf 6_c}$) 
can exhibit color-flavor locking, and thus breaks 
chiral symmetry. This possibility
has been investigated in \cite{ABR} where the sextet contribution
was observed to vanish for sufficiently high strange quark mass. 
One-gluon exchange interactions relevant at asymptotically high densities  
do not support the ${\bf 6_c}$ part of (\ref{delcond}). The
same can be observed using effective instanton induced interactions.  
In contrast to the three-flavor theory the ${\bf 6_c}$-condensate
is not induced by the color-antitriplet condensate since they 
leave different symmetries invariant.
There are no indications that the two-flavor (infinite strange quark) 
limit of (\ref{di3CFL}) acquires a sextet contribution and one is led
to conclude that no color-flavor locking, and consequently no chiral
symmetry breaking, occurs in a two-flavor theory with diquark
condensates only\footnote{The quarks of the third color not involved
in $\Delta^{(2)}_{\bf \bar{3}_c}$ can form a condensate.
In ref.\ \cite{ARW} a small nonzero gap from quarks of 
the third color was found to be a few orders of magnitude 
smaller than the antitriplet gap using effective instanton
interactions. This gap can not break the residual gauge
symmetry since it only involves quarks of the third color.}.\\

{\it Integer charges and quark description of nuclear matter.\ }
Ordinary electromagnetism is broken in 
presence of the diquark condensate $\Delta^{(2)}$ or the
adjoint chiral condensate $\chi^{(2)}$ but, as in the three flavor
case, the condensates are invariant under the combination 
$\tilde{Q}=Q- \frac{1}{2}\lambda_3-\frac{1}{2 \sqrt{3}} \lambda_8$
of electric charge $Q$ and abelian color charge. In particular,
all dressed excitations acquire integer charges. Baryon number 
is broken in presence of a diquark condensate but the  
color-antitriplet diquark condensate $\Delta^{(2)}_{\bf \bar{3}_c}$
is invariant under a modified baryon number 
$\tilde{B}=B-\frac{1}{\sqrt{3}}\lambda_8$ \cite{ARW}. 

Neither the adjoint chiral condensate nor the antitriplet
diquark condensate alone break color completely. In particular,
chiral symmetry is unbroken in presence of the antitriplet
diquark condensate alone. It is striking to observe that in a 
phase with simultaneous condensation in both channels color is 
broken completely, all gluons acquire a gap, and chiral symmetry is broken
(see also \cite{BWW}). 
In this case the integer charged degrees of freedom for the most part match 
the expected features of ordinary nuclear matter! This complements 
the proposal set forward in \cite{SWCONT,SWCONT2} 
of a quark description of nuclear matter phases.
There are only two degrees of freedom, corresponding to the quarks of 
the third color, which carry baryon number $\tilde{B}=1$. The dressed
up quark of the third color carries electric charge 
$\tilde{Q}=+1$ and the corresponding down quark is electrically neutral.
These are the only low energy baryonic degrees of freedom 
and share the properties of the proton and the neutron, respectively.
The remaining dressed triplet and singlet acquire a large gap and
are heavy, since only the quarks of the first two colors are involved
in the diquark gap. They carry integer charges $\tilde{Q}=1,0,-1$ for 
the triplet and $\tilde{Q}=0$ for the singlet and zero baryon number 
$\tilde{B}$. There is also a triplet of collective modes associated 
with chiral symmetry breaking. The broken symmetry generators
are given by the axial charges and the massless bosons match the 
quantum numbers of the three pions. 

There is an octet of massive gluons with quantum numbers of  
the octet of vector mesons. It is surprising at first sight 
that in a pure two-flavor theory there seem to be states
known from a three-flavor theory, which is 
a consequence of the complete breaking of color. We note that
the diquark condensate contributes to the mass of the
$K^*$- and $\omega$-meson but not to the mass of the $\rho$-meson.
The situation is similar to the fermionic sector
discussed above where an additional heavy triplet and singlet
appeared. The low energy theory should contain the expected
two-flavor degrees of freedom. We also note that 
the vector meson $\sim \lambda_8$, which is a singlet under isospin, 
plays the role of the $\omega$ in the two flavor theory. \\ 

We emphasize that there is no guarantee that condensation occurs
in the adjoint quark-antiquark channel. Only for sufficiently strong 
coupling condensation can occur in this channel. Ref.\ \cite{CWINST}
points out that in three-flavor QCD the instanton induced axial anomaly 
results in a cubic term in the effective potential for scalar 
color-singlet and octet quark-antiquark 
states. This induces a nonzero color-octet and singlet quark-antiquark 
condensate in the vacuum \cite{CWINST}\footnote{In contrast, 
for standard mean field treatments as in \cite{klevansky} the 
instanton interaction is attractive in the Hartree approximation. 
The Fock contribution gives a large cancellation and renders the 
interaction repulsive. One-gluon exchange induced interactions, relevant at 
asymptotically high densities, are repulsive in the adjoint 
quark-antiquark channel.}. A similar mechanism for two flavors is discussed in
\cite{BWINST}.   

In the following we will show in a simple quark model that for sufficiently
strong attractive interaction in the adjoint quark-antiquark channel
simultaneous condensation occurs for not too high densities. 
The main outcome is described in fig.\ \ref{2sccfl}. 
The result is encouraging. 
The equivalent calculation looking for condensation in the standard 
chiral condensate and the color-antitriplet diquark channel
found no simultaneous condensation \cite{BR,CD}.

\section{A simple model for two-flavor color-flavor locking}

\hspace*{0.7cm}
{\bf Color-octet and antitriplet channels.\ } We consider a class of 
fermionic models similar to the Nambu--Jona-Lasinio model \cite{NJL} for 
QCD where quarks interact via effective 
four-fermion interactions. 
The model reflects the two-flavor chiral symmetry of
QCD as an $O(4) \sim SU(2)_L \times SU(2)_R$ symmetry. 
Color is realized as a global symmetry. We will only use this model here
to describe the qualitative behavior of quarks in the high density phase
and do not apply it to the vacuum. 
In the presence of a nonvanishing chemical potential $\mu$ 
for (net) quark number, the quadratic part $S_{0}$
of the Euclidean action $S=S_0+S_{\rm INT}$ in momentum 
space is\footnote{We use the conventions employed in ref.\ \cite{BR}.}
\beq
S_{0}= \int \frac{d^4 q}{(2\pi)^4}
\bar{\psi}_a^{\alpha}(q) \left(
\gamma^{\nu}q_{\nu}+i m+ i \gamma^{0} \mu\right)\psi_a^{\alpha}(q) \, 
\label{s0}
\eeq
with an average current quark mass $m$. The interaction part
is taken to contain an interaction in the color-antitriplet
diquark channel and in the color-octet quark-antiquark channel,  
$S_{\rm INT}=S_{\rm INT}^{\bf (\bar{3})}+S_{\rm INT}^{\bf (8)}$
with
\bea
S_{\rm INT}^{(\bar{\bf{3}}_c)} &=&  
- G_{\Delta} \int \frac{d^4 p}{(2\pi)^4} \Big\{
O^{*\, j}_{(\Delta)}[\bar{\psi};p]
  O^{j}_{(\Delta)}[\psi;p] \Big\}\,\, ,
\nonumber\\[0.3cm]
S_{\rm INT}^{(\bf{8}_c)} &=& - G_{\chi} \int
\frac{d^4 p}{(2\pi)^4}\Big\{  O_{(\delta_8)}^{si}[\psi,\bar{\psi};-p] 
 O_{(\delta_8)}^{si}[\psi,\bar{\psi};p]   
\nnn &&   
+ O_{(\eta_8)}^i[\psi,\bar{\psi};-p] 
 O_{(\eta_8)}^i[\psi,\bar{\psi};p]
 \Big\}  \,\, .
\eea
Here we have defined the fermion bilinear $O_{(\Delta)}^j$ $(j=1,2,3)$
carrying the quantum numbers of a Lorentz-scalar color-antitriplet diquark,
the scalar color-octet flavor-triplet bilinear $O_{(\delta_8)}^{s i}$
($s=1,2,3$; $i=1,\ldots,8$) and the corresponding 
pseudoscalar flavor-singlet quark-antiquark
bilinear $O_{(\eta_8)}^i$, respectively, 
\bea
O_{(\Delta)}^j &=& - (\psi^T)^{\alpha}_{a} C \gamma_5 (\tau_2)_{ab}
\big( \lambda^{(A)}_j \big)^{\beta\alpha}  \psi^{\beta}_b \,\, ,
\nonumber \\
O_{(\delta_8)}^{s i}
&=& - i \bar{\psi}^{\alpha}_{a} \left(\tau_s \right)_{ab} 
\left(\lambda_i \right)^{\beta\alpha} \psi^{\beta}_b \,\, ,
\nonumber \\
O_{(\eta_8)}^i &=& - \bar{\psi}^{\alpha}_{a} \gamma_5 
\left(\lambda_i \right)^{\beta\alpha} \psi^{\beta}_b \,\, . 
\eea
We will discuss 
a mean field solution for the low free-energy modes of the model 
and study condensation in the adjoint chiral channel (\ref{chicond})
and the color-antitriplet diquark channel of (\ref{delcond})\footnote{We 
will not discuss possible condensation in
the pseudoscalar channel.}.
We use the model at nonzero density with 
quark chemical potentials $\mu \gtrsim 350$ MeV
as a reasonable lower bound motivated by the expected onset of nonzero
density around $\mu \simeq m_{\rm Nucleon}/3$.  
We regulate the model by an effective cutoff scale of about  
$1/2 - 1$ GeV. 
For the present investigation the results are rather insensitive to
the details and to be explicit we  
implement a cutoff in three-momentum space of $\Lambda=0.7$ GeV.
We fix the coupling in the diquark channel, $G_{\Delta}$,
to obtain color superconducting gaps of typical QCD scales of
${\cal O} (100 {\rm\ MeV})$ consistent with estimates in 
effective models and perturbative estimates \cite{review}.
(We use here a diquark gap $\Delta=125$ MeV at $\mu=500$ MeV 
for $G_{\chi}=0$.) 
We will study the phase structure of the model as a function 
of quark number chemical potential $\mu$ ($\equiv \mu_{Baryon}/3$) 
and the coupling in the octet channel, $G_{\chi}$.

Based on earlier studies in \cite{BR} we
expect the standard chiral condensate to be small in presence of a 
nonzero diquark condensate. Though a strong competition
between chiral and diquark condensate was observed we note that
a small chiral condensate will be present in a phase with
simultaneous adjoint quark-antiquark and quark-quark condensates.
This is due to the fact that in the latter phase no new symmetries
are broken by a nonvanishing chiral condensate. We verified this
explicitly  by adding an attractive 
interaction to the model in the color-singlet quark-antiquark channel.
The condensate turned out to be very small for sufficiently high density
in the presence of simultaneous adjoint chiral and diquark condensates which
we will discuss below, and we neglect the 
standard chiral condensate in the following. \\

{\bf Thermodynamic potential.\ }
The mean field effective potential $\Omega(\chi,\Delta; \mu)$ as
a function of the adjoint chiral gap parameter $\chi$ and the
diquark gap parameter $\Delta$ and quark chemical potential $\mu$ 
can be obtained along the lines of ref.\ \cite{BR}. At its extrema, 
the potential is related to the energy density $\epsilon$, 
the standard quark number density $n$ 
and the pressure $P$ by
\beq
\Omega(\chi_0,\Delta_0;\mu) = \epsilon - \mu n = -P\ ,
\eeq
where the extremum condition
\begin{equation}
\frac{\partial \Omega}{\partial\chi}\Bigg|_{\chi=\chi_0;
\Delta=\Delta_0}=\frac{\partial \Omega}{\partial\Delta}\Bigg|_{\chi=\chi_0;
\Delta=\Delta_0} =0\ 
\label{gapeqs}
\end{equation}
leads to coupled gap equations for $\chi_0$ and $\Delta_0$. They are 
related to the adjoint chiral condensate
(\ref{chicond}) and the condensate of Cooper pairs (\ref{delcond}) by 
($a,b,\alpha,\beta=1,2$)
\bea
\chi_0 &=& 2 G_{\chi} \Big\langle \bar{\psi}^{\alpha}_a 
\Big(2 \delta_a^\alpha \delta_b^\beta - \delta_{ab} \delta^{\alpha\beta}\Big)
\psi^{\beta}_b \Big\rangle  \,\, ,
\\[0.2cm]
\Delta_0 &=& 2 G_{\Delta} \Big\langle {\psi}^{\alpha}_a 
\Big(\delta_a^\alpha \delta_b^\beta 
- \delta_{a}^\beta \delta_b^{\alpha} \Big)
\psi^{\beta}_b \Big\rangle \,\, . 
\eea
The effective potential reads
\bea \lefteqn{
\Omega(\chi,\Delta;\mu) = \frac{3}{4 G_{\chi}} \chi^2 
+ \frac{1}{4 G_{\Delta}} \Delta^2 }
\nnn 
&-& \int_0^{\Lambda} 
\frac{q^2 dq}{2 \pi^2} \Biggr\{ \,  
3 (E_{\chi}^{-}-\mu) \sqrt{1+ \Delta^2/(E_{\chi}^{-}-\mu)^2}\,
\Big( \Theta(E_{\chi}^{-}-\mu) - \Theta(\mu-E_{\chi}^{-}) \Big)
\nnn
&&\hspace*{1.6cm}
+ 3 (E_{\chi}^{-}+\mu) \sqrt{1+ \Delta^2/(E_{\chi}^{-}+\mu)^2} 
\nnn
&&\hspace*{1.6cm}
+ (E_{\chi}^{+}-\mu) \sqrt{1+ \Delta^2/(E_{\chi}^{+}-\mu)^2}\,
\Big( \Theta(E_{\chi}^{+}-\mu) - \Theta(\mu-E_{\chi}^{+}) \Big)
\nnn
&&\hspace*{1.6cm}
+ (E_{\chi}^{+}+\mu) \sqrt{1+ \Delta^2/(E_{\chi}^{+}+\mu)^2} 
\nnn
&&\hspace*{1.6cm}
+ 4 \left( E\, \Theta(E-\mu) + \mu\, \Theta(\mu-E) \right)
\,\,+\,\, {\rm constant} \Biggl\}
\label{potential}
\eea
with $q=|\vec{q}|$, a field-independent $E(q) = \sqrt{{q\, }^2+m^2}$
and     
\beq
E_{\chi}^{-}(q) = \sqrt{{q\, }^2+(m - \chi)^2} \quad, \qquad
E_{\chi}^{+}(q) = \sqrt{{q\, }^2+(m+ 3 \chi )^2} \, .
\eeq
The constant in (\ref{potential}) does not depend on $\mu$ 
and is chosen such that the pressure of the physical vacuum is 
zero.

As $\mu$ or $G_{\chi}$ changes the potential can have several
local minima in the $(\chi,\Delta)$--plane. The lowest minimum
describes the lowest free energy state and is favored. Apart from the
trivial vacuum with no condensation, in principle one can find
three types of global minima. One corresponds to a ground state 
characterized by a nonzero diquark gap $\Delta \not = 0$ with $\chi=0$. 
This is the two-flavor color superconducting state (2SC) 
\cite{Barrois,BailinLove,ARW,Rapp,BR}. 
The other possibility is a minimum 
with no diquark condensate but a nonzero mass gap in the adjoint 
quark-antiquark channel. The third possibility is a ground state   
characterized by a simultaneous diquark and
adjoint chiral condensate. This state exhibits two-flavor color-flavor 
locking (2CFL) where both chiral and color symmetries are broken
\begin{figure}[bht]
\begin{center}
\epsfig{file=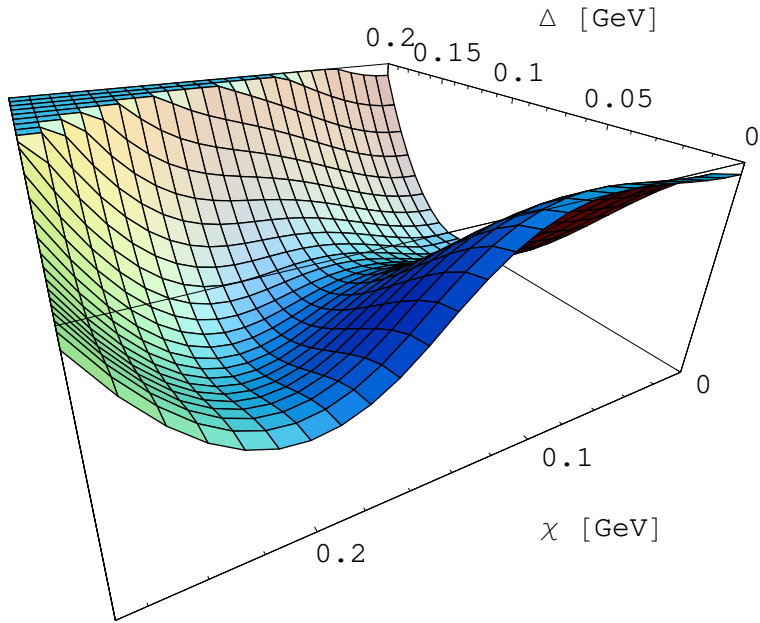,width=10.cm
,bbllx=110,bblly=440,bburx=440,bbury=720}
\end{center}
\caption{Effective potential at a density where 
2CFL and 2SC phases coexist.} 
\label{potcd2}
\end{figure}
completely. Fig.\ \ref{potcd2} shows an
example where the potential has two degenerate minima corresponding to
a first-order phase transition at which two phases have equal
pressure and can coexist ($\mu=475$ MeV, $G_{\chi}/G_{\Delta}=2.78$, 
$m=0$)\footnote{The results discussed in this work are  
rather insensitive to a nonzero small quark mass $m$.}. 
One minimum occurs at $\chi=187$ MeV,
$\Delta=88$ MeV and corresponds to the 2CFL phase. The other
minimum occurs at $\chi=0$, $\Delta=120$ MeV and corresponds
to the 2SC phase. For smaller $\mu$ the color-flavor locked
state becomes the global minimum. We will show in the following 
that, for sufficiently large coupling in the adjoint chiral channel, 
the generic picture which emerges from the model is that a 
2CFL phase appears once $\mu$ is lowered below a critical value. \\

\section{Simultaneous condensation at high density}

\hspace*{0.7cm}
{\bf Two-flavor color-flavor locking.\ }
\unitlength1.0cm
\begin{figure}[t]
\begin{picture}(1.0,1.0) 
\put(9.3,-4.3){\bf\large ${2SC}$}
\put(5.5,-2.5){\bf\large ${2CFL}$}
\end{picture}
\begin{center}
\epsfig{file=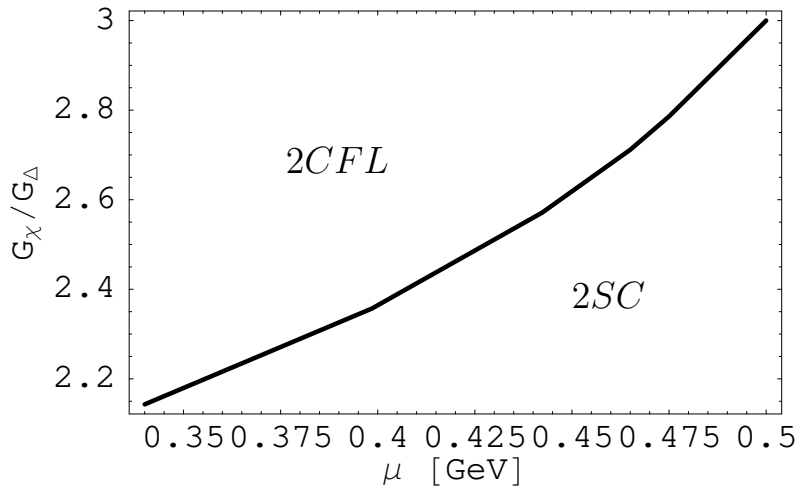,width=9.5cm
,bbllx=100,bblly=550,bburx=366,bbury=712}
\end{center}
\caption{Phase structure as a function of quark chemical
potential $\mu$ and couplings $G_{\chi}/G_{\Delta}$.
Shown is the coexistence line between the 2CFL and the
2SC phase. 
}
\label{2sccfl}
\end{figure}
To observe a few general properties of the phase structure
we consider first the possibility of a state with no diquark
condensate, $\Delta=0$, and $\chi \not = 0$.   
At sufficiently high density or chemical potential $\mu$ such a
state can not occur since more and more low momentum contributions, 
with momenta $q$ for which $E_{\chi}^\pm(q)<\mu$, are absent 
in the gap equation for $\chi$ as 
$\mu$ increases (cf.\ eqs.\ (\ref{potential}), (\ref{gapeqs})). 
As a consequence 
$\chi$ vanishes for large enough $\mu$. In contrast,
the BCS mechanism is operative for quark-quark
Cooper pairs and ensures a nonzero diquark gap even at 
asymptotically high density. 

A state with $\chi \not = 0$, either with zero or nonzero diquark gap, 
can occur at not too high densities. In fig.\ \ref{2sccfl} we plot
the phase diagram of the model (\ref{potential}) as a function of $\mu$ 
and $G_{\chi}/G_{\Delta}$. We cover a range for the chemical potential
from $\mu \gtrsim 350$ MeV to $\mu \lesssim 500$ MeV. For even higher
values of $\mu$ the two-flavor discussion should become unappropriate
since before one may expect strange quarks to be present. 
We observe that for a sufficiently strong coupling in the
adjoint quark-antiquark channel simultaneous condensation 
happens always for small enough density or $\mu$ in this range. 
This is nontrivial, in particular, since simultaneous
condensation in the diquark and the standard color-singlet 
quark-antiquark channel could not be observed at any
density \cite{BR,CD}. Though this opens striking possibilities
for a quark description of ordinary nuclear matter as
pointed out above, we emphasize that the possibility
of quark-antiquark condensation is constrained by the strength
of the coupling. For too small $G_{\chi}$ the 2SC phase
is realized.   
\\

\begin{figure}[h]
\begin{picture}(0.1,0.1) 
\put(9,-2){\bf\large ${2SC}$}
\put(5,-3.5){\bf\large ${2CFL}$}
\end{picture}
\begin{center}
\epsfig{file=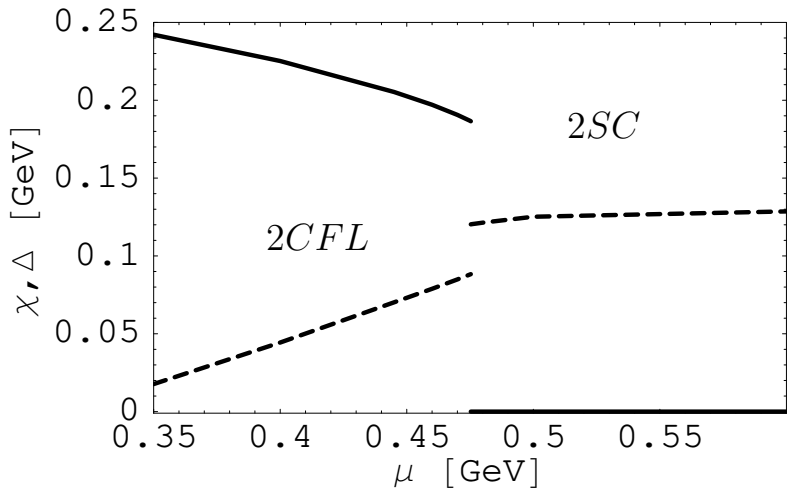,width=10.cm
,bbllx=100,bblly=550,bburx=366,bbury=712}
\end{center}
\caption{Adjoint quark-antiquark (solid) and diquark (dashed) gaps
as a function of $\mu$ for the same couplings as in fig.\ \ref{potential}. 
One observes an unlocking transition from the 2CFL phase to the 2SC
phase at a critical $\mu$. The 2SC phase may not be realized if
strange quarks are taken into account.     
}
\label{chidel}
\end{figure}

{\bf Unlocking transition and onset of the strange quark.\ }
The phase diagram \ref{2sccfl} shows an unlocking transition 
from the two-flavor color-flavor locked phase to the 2SC phase as $\mu$ 
increases. This phase transition is first order. Fig.\
\ref{chidel} shows the behavior of $\chi$ (solid line)
and $\Delta$ (dashed line) as a function of $\mu$
for the same couplings as in fig.\ \ref{potential}.
At a critical value of $\mu=475$ MeV one observes that the
adjoint chiral condensate vanishes discontinuously while the
diquark gap jumps to a higher value. 

If the 2CFL phase is realized in nature at low enough densities then
the situation as the density increases depends strongly on the value 
of the strange quark mass. Two qualitative different situations
can occur. For sufficiently large strange quark mass the 2CFL phase
would terminate in the 2SC phase, restoring chiral symmetry
and breaking color only partially to a residual $SU(2)$.
As the density is further increased one finally enters
the ``standard'' three-flavor color-flavor locked
(CFL) phase with a first order transition \cite{ABR},
again breaking chiral symmetry and color completely as in
the 2CFL phase.   

In contrast,
for sufficiently small strange quark mass there
can be a transition directly from the 2CFL to the three-flavor 
CFL phase. The additional
symmetries broken during this transition are baryon number
and strangeness. In the present model the 2CFL phase is no 
superfluid. However, if the quarks
of the third color condense this would 
break baryon number. This is likely since the BCS mechanism
guarantees a gap in presence of an arbitrarily small attractive
interaction. In ref.\ \cite{ARW} such a gap was observed
in a similar model which was a few orders of magnitude 
smaller than the antitriplet gap. Since the
quarks of the third color carry the same quantum numbers
as the proton and the neutron the possible 
channels for condensation correspond precisely to the known 
possibilities for pairing in nuclear matter. This fact is valid
for the 2SC phase as well and has been
pointed out in \cite{SWCONT2}. We conclude that 
color and chiral symmetry would be broken at all nonzero
densities in the latter case, with an equivalent description of 
confinement in the Higgs picture and qualitative properties which 
match the expected features of nuclear as well as hypernuclear matter.

\vspace*{1cm}

\noindent
{\bf Acknowledgement}\\

I would like to thank Christof Wetterich and Uwe-Jens Wiese for collaboration 
on related work in ref.\ \cite{BWW} and Mark Alford for helpful 
discussions.

\end{document}